\input{aipcheck}

\documentclass[
    ,final            
  ]
  {aipproc}
\layoutstyle{6x9}

\begin{document}

\title{Simulations of relativistic collisionless shocks: shock structure and particle acceleration }

\classification{}
\keywords      {}

\author{Anatoly Spitkovsky}{
  address={Kavli Institute for Particle Astrophysics and Cosmology,\\
Stanford University, PO Box 20450, MS 29, Stanford, CA 94309}
}

\begin{abstract}

We discuss 3D simulations of relativistic collisionless shocks in
electron-positron pair plasmas using the particle-in-cell (PIC)
method. The shock structure is mainly controlled by the shock's
magnetization ("sigma" parameter). We demonstrate how the
structure of the shock varies as a function of sigma for 
perpendicular shocks. At low magnetizations the shock is
mediated mainly by the Weibel instability which generates transient
magnetic fields that can exceed the initial field. At larger
magnetizations the shock is dominated by magnetic reflections. We
demonstrate where the transition occurs and argue that it is
impossible to have very low magnetization collisionless shocks in
nature (in more than one spatial dimension). We further discuss the
acceleration properties of these shocks, and show that higher
magnetization perpendicular shocks do not efficiently accelerate
nonthermal particles in 3D. Among other astrophysical applications,
this may pose a restriction on the structure and composition of gamma-ray
bursts and pulsar wind outflows.

\end{abstract}

\maketitle


\section{Introduction}
Relativistic collisionless shocks are commonly encountered in astrophysics, mostly where relativistic flows are converted into observable radiation. They are thought to occur at the termination of pulsar winds in pulsar wind nebulae (PWN), in AGN jets and in gamma-ray bursts (GRB). The flows in these objects range in relativistic factor ($\gamma \sim 2-10$ in AGNs, $10-100$ in GRBs, and upto $10^6$ in PWN), magnetization (ratio of magnetic to kinetic energy $\sigma \sim 10^{-3}-1$ in PWNe, unknown in GRBs and AGNs, models range from small to very large magnetization), and in composition (unknown in all three; models include $e^\pm$ pairs, pairs+ions to electron-ion plasmas). Despite these differences, collisionless shocks are expected to have common features, inferred from observations. Such shocks are \emph{expected} to provide effective viscosity to mediate MHD jump conditions without collisions, as well as  generate magnetic fields, and efficiently accelerate nonthermal particles with powerlaw spectra. We aim to evaluate these expectations by constructing numerical ab-initio 3D models of collisionless shocks in electron-positron pair plasma and studying the dependence of shocks structure on the properties of the upstream flow (Spitkovsky and Arons, in preparation). 

\section{Simulation setup}
The processes leading to the formation of collisionless shocks involve dynamics on the fundamental plasma scale, therefore to model such shocks we require a plasma simulation code. We use particle-in-cell method (PIC) (e.g., \cite{birdsall}) for electromagnetic plasma simulation. We represent plasma as a collection of macroparticles and solve inhomogeneous Maxwell equations with currents provided by motion of macroparticles. The motion of the particles in self-consistent fields is computed using Lorentz force. We have extensively modified the publicly available code TRISTAN \cite{buneman}, which is a 3D electromagnetic PIC code in Cartesian coordinates. We improved the behavior of the code in the ultrarelativistic regime to avoid numerical grid-Cerenkov radiation, added filtering of current data to improve noise properties, and fully parallelized the code to efficiently run on hundreds of processors. 

In order to set up a collisionless shock we can collide two plasma shells moving at each other. This will initiate two shock waves propagating into the upstream. For most of the simulations we chose to save half the size of the simulation by reflecting one plasma stream off a wall. This sends a reverse shock into the domain, as if another stream is present outside box. Initially the stream is charge neutral and cold, moving at $\gamma=15$. We mainly use equal mass macroparticles of both signs of  charge (pairs), with limited experiments with different mass ratios.  In order to be stable a PIC code requires the plasma oscillation frequency $\omega_p=(4\pi n e^2/\gamma m)^{1/2}$ and skindepth $c/\omega_p$ be resolved. We usually have $c/\omega_p=10$ cells in the upstream region and simulate domains upto $200\times  40 \times 40$ skindepth on the side, or $2000\times 400 \times 400$ cells with several billion particles. 3D PIC codes are currently also used by other groups to study shocks, notably by \cite{silva, nishikawa, nordlund}. All simulations agree on the general physical processes involved, but groups differ in the  setup of the simulations and the length and extent of the runs. We tried to run simulations large enough to resolve the full structure of the shock, and investigated shock properties as a function of the upstream magnetization parameter $\sigma=$magnetic energy/kinetic energy$=\omega_c^2/\omega_p^2=B^2/(4 \pi n \gamma m c^2)$ in the range from 0 to 10. 

\section{Shocks with varying magnetization}
\begin{figure} 
\unitlength = 0.0011\textwidth
\hspace{1\unitlength}
\begin{picture}(400,400)(0,0)
\put(-20,25){\makebox(0,0){ (a)}}
\includegraphics[scale=.32]{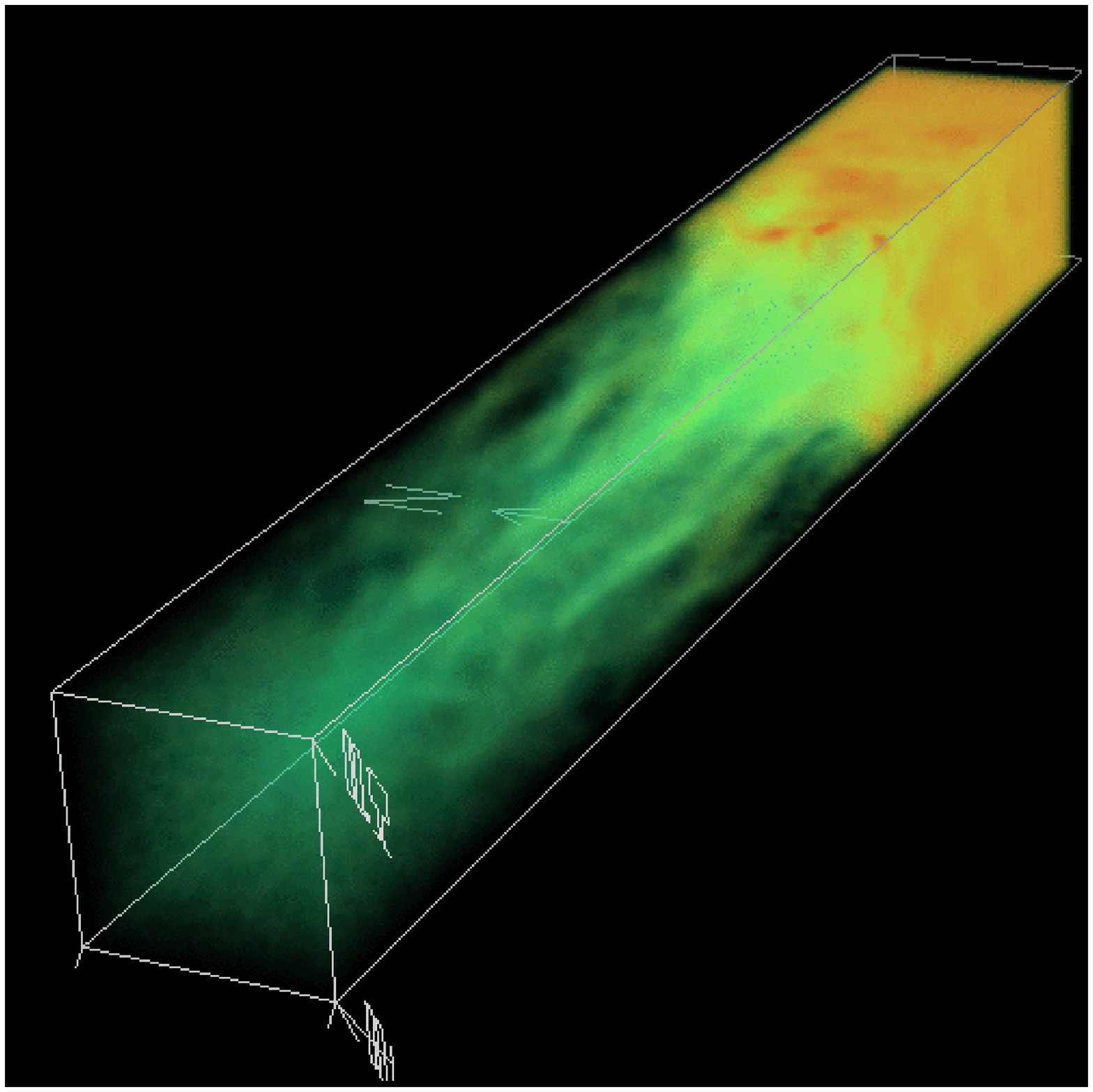}
\end{picture}
\hspace{1\unitlength}
\begin{picture}(400,400)(0,0)
\put(-20,25){\makebox(0,0){ (b)}}
\includegraphics[scale=.32]{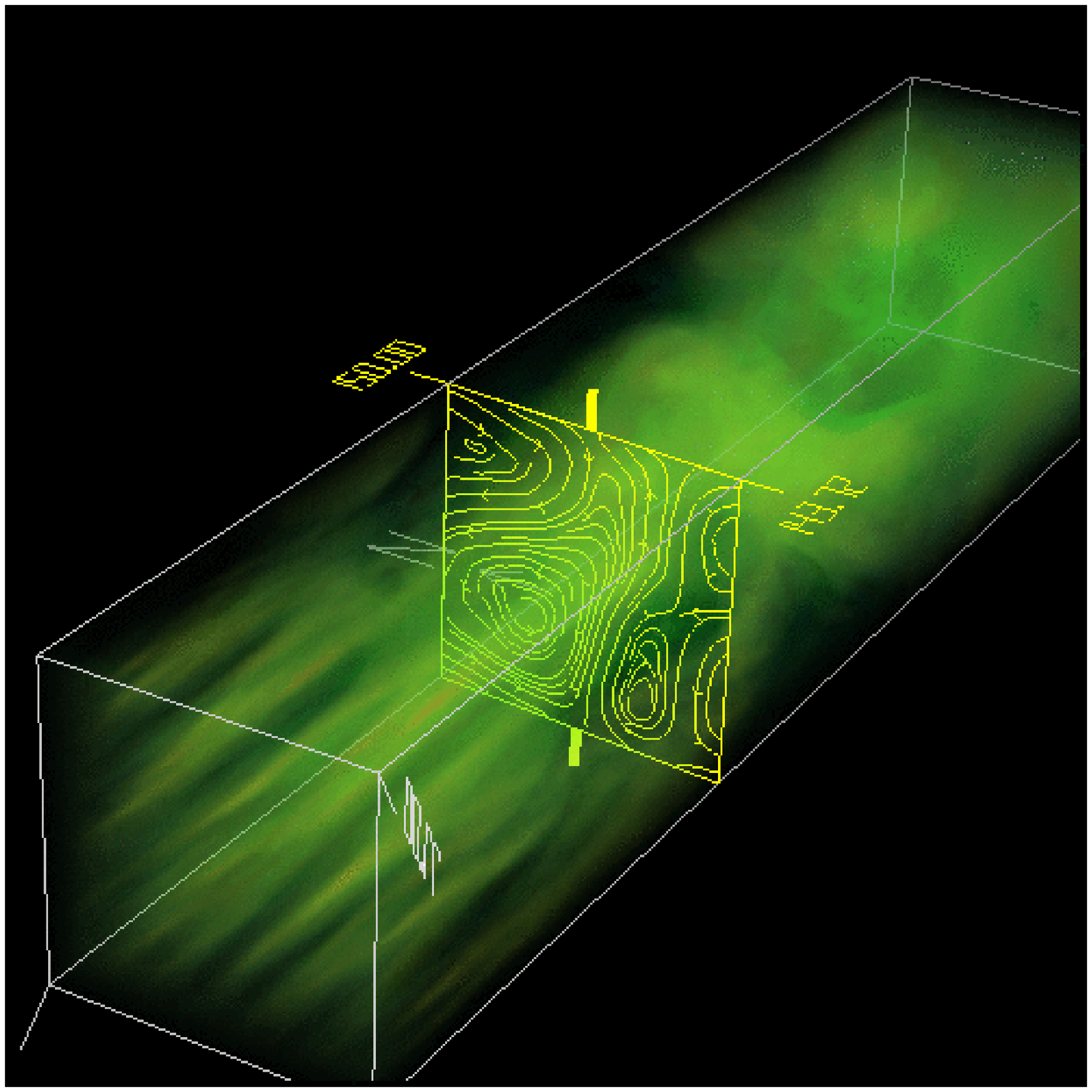} 
\end{picture}
\caption{a) Filamentary structure of density in an unmagnetized shock. b) Generation of magnetic field around current filaments in the shock.
}
\label{fig3dsig0}
\end{figure}

\begin{figure} 
\unitlength = 0.0011\textwidth
\hspace{1\unitlength}
\begin{picture}(400,400)(40,0)
\includegraphics[scale=.4]{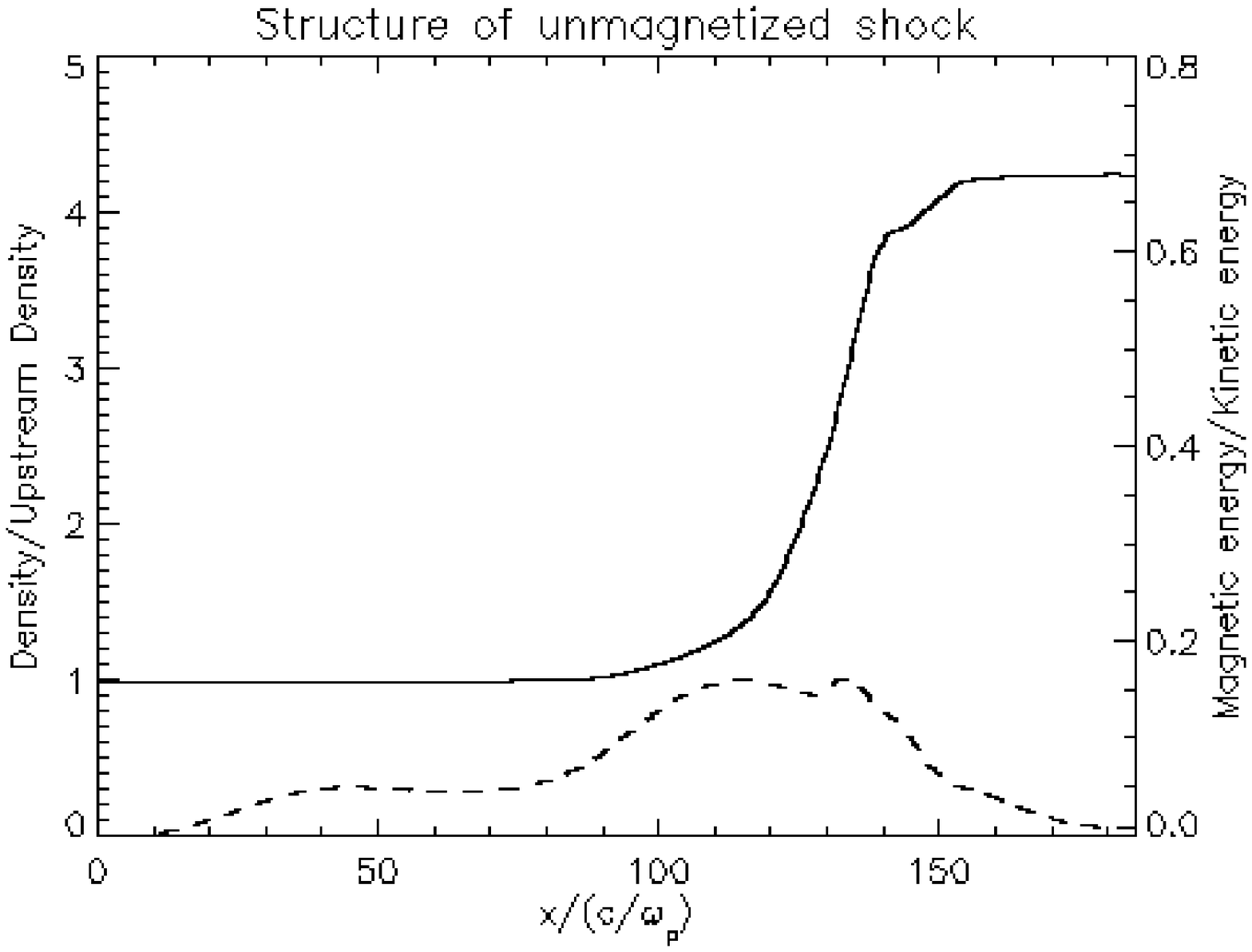}
\put(-420,25){\makebox(0,0){(a)}}
\end{picture}
\hspace{1\unitlength}
\begin{picture}(400,400)(0,0)
\includegraphics[scale=.4]{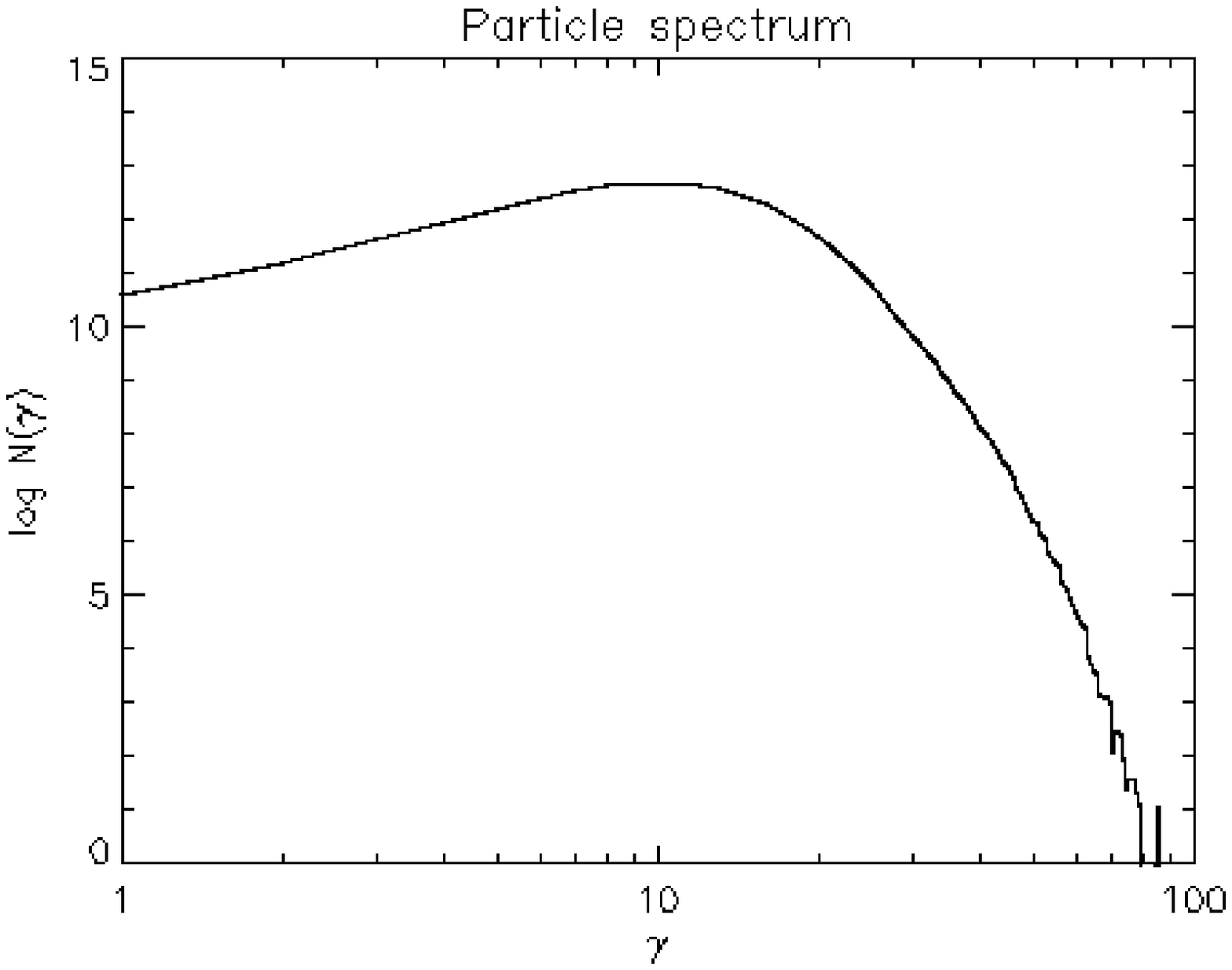} 
\end{picture}
\put(-380,25){\makebox(0,0){ (b)}}
\caption{a) Density structure through the unmagnetized shock (solid line, left axis) and magnetic energy normalized by the upstream kinetic energy (dashed line, right axis). b) Downstream particle spectrum.
}
\label{unmagstruct}
\end{figure}

We first consider unmagnetized $\sigma=0$ pair shocks. As particles reflect off the wall, a counterstreaming distribution is set up between the reflected and upstream particles. This distribution is unstable to Weibel instability \cite{Weibel59}. This streaming instability is electromagnetic in nature, as opposed to electrostatic streaming instabilities common in nonrelativistic shocks. The physical picture can be understood by considering a small perturbation of magnetic field in the plane perpendicular to the direction of motion \cite{MedvedevLoeb99}: the magnetic field deflects positive and negative charges in opposite directions transverse to the flow. This leads to separation of charge and current into filaments. Even though initially the flow is current free, it is quickly partitioned into filaments of oppositely flowing currents, which locally amplify the magnetic field that led to the creation of filaments. This leads to a runaway which forms filaments not only in current but also in particle density. Figure \ref{fig3dsig0}a shows 3D density distribution through such a shock. The initial scale of the filaments is the plasma skindepth $c/\omega_p$, and the growth rate is $\sim 10/\omega_p$, so the instability grows on astrophysically microscopic scales. The magnetic field is initially created on the plasma scale in loops in the plane transverse to the direction of motion, as shown in figure \ref{fig3dsig0}b. As more plasma is segregated into current filaments they reach Alfven critical current at which point the currents are unstable to self-pinching. Particles in the plasma get scattered by the self-generated magnetic field and the average velocity in the flow direction decreases. Correspondingly, plasma density starts to increase, and approaches the density of the Rankine-Hugoniot jump condition. In Figure \ref{unmagstruct}a we show the density structure through an unmagnetized shock. Our simulations are done in the downstream frame, and the shock is moving through the domain. The jump condition in this frame is $n_2/n_1=\Gamma/(\Gamma-1)+O(\sigma)$, where $\Gamma$ is the adiabatic index, which is equal to $4/3$ for the relativistic plasma in the simulation. Consequently we get a density jump of roughly $4$ in this case, and the shock moves at $c(\Gamma-1)=c/3$. The magnetic field energy in the box, shown with dashed line in fig. \ref{unmagstruct}a, shows a peak in the middle of the shock at about $16\%$ of the upstream kinetic energy in the flow, and then decays to less than 1$\%$ in effectively downstream region where density saturates. Our simulations do not presently allow to measure the field evolution below this level, as it gets dominated by noise. A simulation with larger number of particles and lower noise is needed to answer whether the magnetic field energy saturates or completely decays. 

A closer examination of fig. \ref{unmagstruct}a shows another peak of magnetic energy in the upstream region. This is the effect of the initialization of shocks in our simulations. When two clouds collide (or, equivalently, a flow is reflected off a wall) the first counterstreaming particles fly through the upstream region virtually unimpeded. The Weibel instability happens behind them and gradually stops the bulk of the counterstreaming flow, but the particles at the head continue to plough through the upstream, forming a precursor. Their density is much smaller than that of the upstream flow, so the rate of growth of Weibel instability is decreased, and they deplete over distances much larger than the thickness of the shock. The upstream peak in fig. \ref{unmagstruct}a actually moves with respect to the shock, because the precursor particles fly essentially at $c$, while the shock moves at $c/3$. Eventually, the precursor dies away, but it leaves behind magnetic field in the upstream region and preheats the upstream plasma which slightly modifies the jump conditions. We are currently working on setting up shocks that are not moving through the grid with the hope of capturing the long-term equilibrium structure of the shock. However, the streaming instability of  low density fast particles in the upstream is interesting for self-generated turbulence needed for particle acceleration. 

The particle spectrum in the downstream of the shock is shown in fig. \ref{unmagstruct}b. We observe a very clear thermalization of the flow, with the resulting distribution being a relativistic Maxwellian with a temperature determined by the upstream flow energy. In our simulations we do not see any significant signature of nonthermal tails in the particle distribution, indicating the absence of Fermi acceleration. Previous reports of nonthermal acceleration\cite{nishikawa} can be explained as the beginnings of thermalization, which naturally creates particles with higher (and lower!) energies than in the upstream flow. 

\begin{figure} 
\unitlength = 0.0011\textwidth
\hspace{1\unitlength}
\begin{picture}(400,400)(0,0)
\put(-20,25){\makebox(0,0){ (a)}}
\includegraphics[scale=.32]{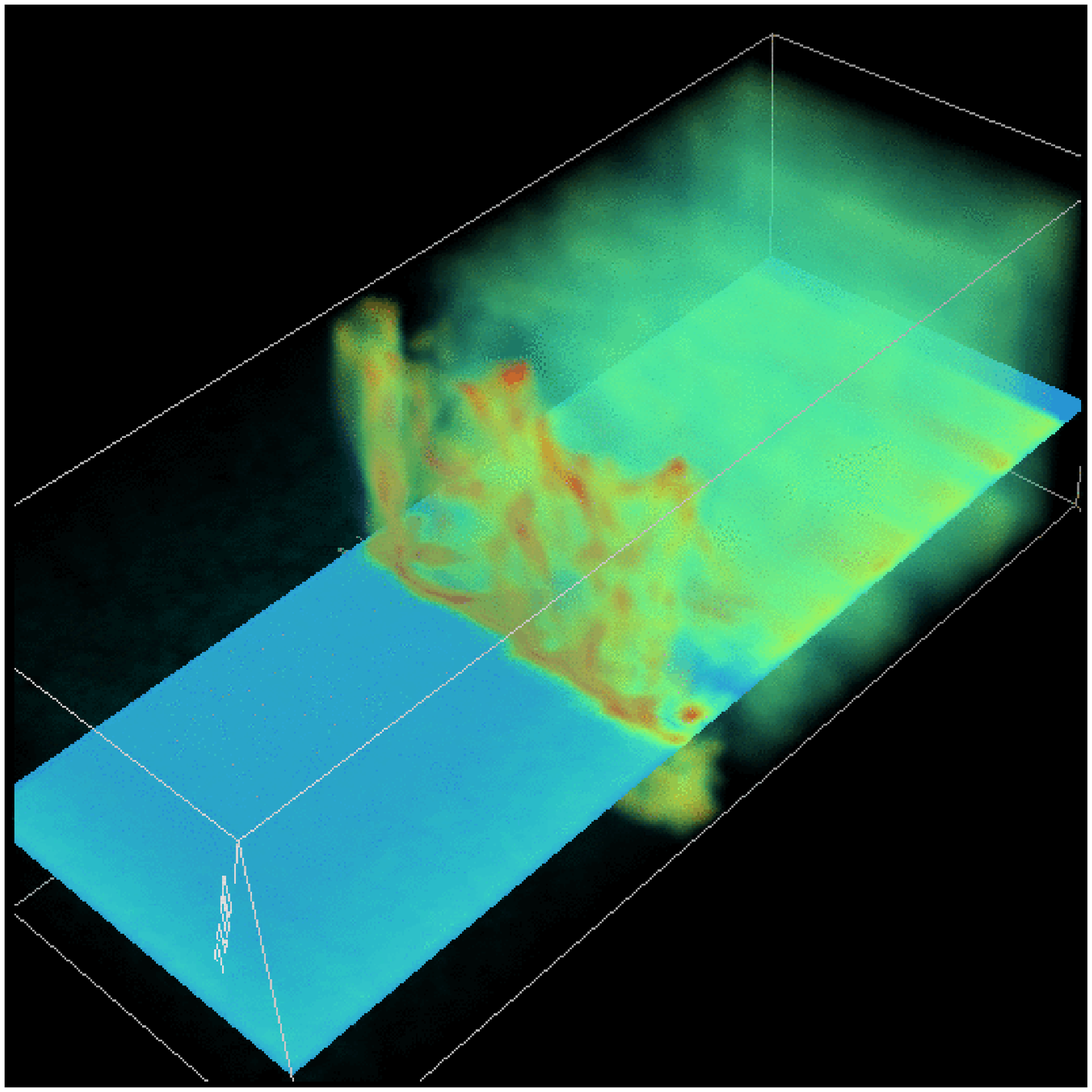}
\end{picture}
\hspace{1\unitlength}
\begin{picture}(400,400)(0,0)
\includegraphics[scale=.4]{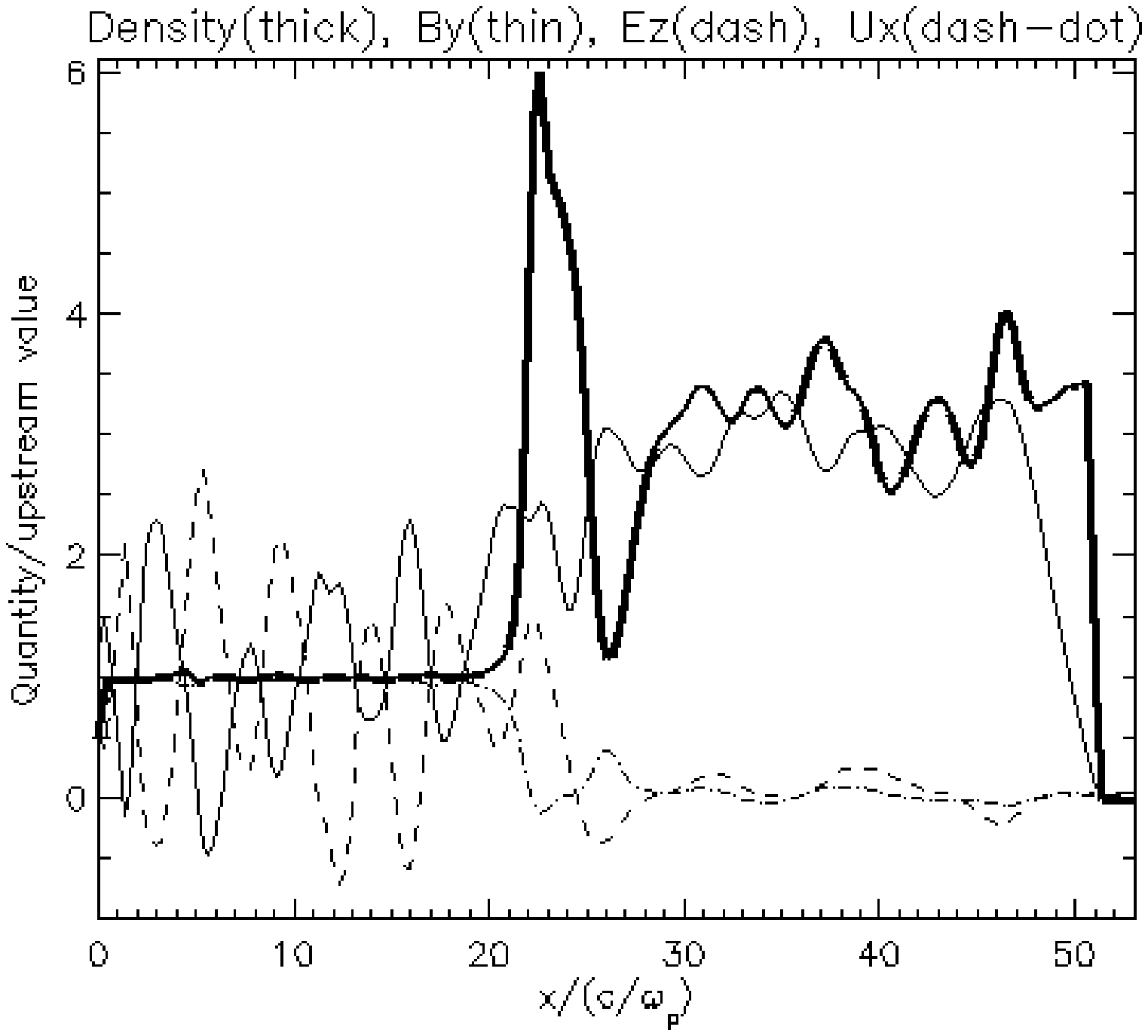} 
\put(-400,25){\makebox(0,0){ (b)}}
\end{picture}

\caption{a) 3D density structure of the magnetized $\sigma=0.1$ shock. Magnetic field is in the shown horizontal plane, perpendicular to the shock normal; b) Averaged density (thick solid line), transverse magnetic field in the horizontal plane $B_y$ (thin solid line), electric field $E_z$ (dashed line), and fluid velocity (dash-dotted line), as a function of distance through the shock, normalized to the value upstream of the shock. 
}
\label{magshock}
\end{figure}

We now turn to magnetized shocks and consider the case with upstream $\sigma=0.1$. The magnetic field is introduced in the horizontal plane, perpendicular to the direction of the flow. The shock density structure in 3D is shown in figure \ref{magshock}a. It is clear that the shock is much sharper and thinner than in the unmagnetized case. When $\sigma=0$ the shock transition, defined by when the density reached a plateau, takes $70 c/\omega_p$, while for $\sigma=.1$ it takes less than $10 c/\omega_p$. The physical nature of the shock is different in the magnetized case. While for $\sigma=0$ the shock was mediated by the Weibel instability, the magnetized shock appears because of magnetic reflection of particles on the compressed magnetic field. Initially all particles follow $E\times B$ motion in the flow direction. The reflected particles (or particles from the opposing cloud), see a wrong sign of the electric field and undergo Larmor gyration. This gyration causes positive and negative particles to go in the opposite directions transverse to the flow, and the associated transverse current increases the magnetic field. The incoming particles now see a jump in the magnetic field and undergo gyration, decelerating and increasing the local density of the flow. In the magnetized regime the shock transition, therefore, happens within a few Larmor radii in the compressed field. There are also bunching instabilities which eventually destroy coherent orbits, but these instabilities are not as germane to the shock mediation as the Weibel  instability is for unmagnetized shocks.

The transversely-averaged quantities as a function of distance through the shock are shown in fig. \ref{magshock}b. The density (thick solid line) shows a sharp compression in the first Larmor orbit and reaches the asymptotic value of roughly 3 times the upstream density. This value can be obtained from $n_2/n_1=\Gamma/(\Gamma-1)+O(\sigma)$ but for a gas with $\Gamma=3/2$. This adiabatic index describes a two-dimensional relativistic gas. The dimensionality is reduced because particles mainly move in the plane perpendicular to the magnetic field. The $\sigma$ correction becomes important for larger $\sigma$ and eventually saturates at $n_2/n_1=2$, or a weak shock, essentially a reflection of a strong electromagnetic wave. The thin solid line in fig. \ref{magshock}b shows the transverse magnetic field in the horizontal plane. It undergoes compression by a factor of $3$. Further fluctuations downstream can reach $\delta B/B\sim 0.5$ (the fluctuations appear much smaller when averaged over transverse dimensions). In front of the shock there is an electromagnetic precursor as seen in the magnetic and electric (dashed line) fields. This is a transverse electromagnetic wave that propagates upstream from the shock. It is generated by the bunching in the coherent Larmor orbits near the front of the shock. The frequency of the wave is comparable to the Larmor gyration frequency. Unlike the unmagnetized case, this precursor is not a transient, and is constantly generated in the shock. The particle spectrum (not shown) in the downstream is very similar to fig. \ref{unmagstruct}b -- a relativistic Maxwellian with no indication of a nonthermal tail. 

Given such difference in shock structure between the unmagnetized and magnetized shocks it is natural to ask what happens for intermediate magnetizations, and where the transition between low and high magnetization actually occurs. We have run a series of shock simulations with $\sigma$ ranging from $0$ to $10$. At the high end, the shocks with $\sigma \gg 0$ look similar to the $\sigma=0.1$ case, but with a decreasing compression ratio. As the Larmor radius becomes smaller in the increased magnetic field, the shock is sharper, and the overshoot in density in the first loop as in fig. \ref{magshock}b is more dramatic. In fact, one can have several of such overshoots before density begins to ramp up. In all simulations the downstream particle spectrum is thermal. As we decrease the magnetization towards $0$, the shock structure begins to change: shock becomes thicker and more filamentary with lower magnetization. While there is no single threshold $\sigma$ for a sharp transition, below a characteristic value of $\sigma=10^{-2}$ the shock is dominated by Weibel instability and is effectively unmagnetized. There are two ways of justifying this value. One is to compare the rate of growth of the Weibel instability with the Larmor gyration time in the background magnetic field. For low magnetic fields, as particles undergo Larmor gyration, they present a counterstreaming distribution which can go Weibel-unstable. If the instability can grow nonlinear within roughly a quarter of the Larmor orbit, the orbit will be disturbed and filamentation will proceed. The self-generated field of the instability dominates the background field for $\sigma < 10^{-2}$. Thus it is impossible in nature to have a very low magnetization shock which would have coherent Larmor orbits. This mistake was made in some 1D simulations \cite{Hoshino03}. 

The absence of any signature of nonthermal acceleration in our simulations is disturbing. There could be several reasons for this: a) simulations require injection of seed high-energy particles and/or longer run time and better statistics; or b) Fermi acceleration does not work, at least in pair plasmas, and ions may be important. We have not ruled out all of these possibilities, but there are some conclusions we can make. We find that there is little possibility of diffusive Fermi acceleration in 3D relativistic magnetized perpendicular pair shocks. There, downstream particles are trapped by the magnetic field and do not return to the shock. Monte-Carlo simulations (e.g. \cite{BednarzOstrowski}) found Fermi acceleration, but required exceptionally high level of post-shock turbulence $\delta B/B > 1$, which were put in by hand. In a self-consistent PIC simulation such level of turbulence is not observed. This does not seem to depend on the resolution or particle number. Injection of high energy beam also led to thermalization, as the particles never scattered upstream of the shock. The weakly-magnetized regime seems to be more promising, however. There, the particles do seem to scatter, and some do return to the shock. However, whether Fermi acceleration can actually take place needs to be investigated with a larger simulation. It would be very interesting if the acceleration properties of the shocks end up depending on the composition of the flow. This would place important constraints on composition of flows in GRBs AGNs and pulsar winds. We hope to answer these questions with further electron-ion simulations.



\bibliographystyle{aipprocl} 



\end{document}